\documentclass[twoside,twocolumn,9pt]{article}
\usepackage{extsizes}
\usepackage[super,sort&compress,comma]{natbib} 
\usepackage[version=3]{mhchem}
\usepackage[left=1.5cm, right=1.5cm, top=1.785cm, bottom=2.0cm]{geometry}
\usepackage{balance}
\usepackage{widetext}
\usepackage{times,mathptmx}
\usepackage{sectsty}
\usepackage{graphicx} 
\usepackage{lastpage}
\usepackage[format=plain,justification=raggedright,singlelinecheck=false,font={stretch=1.125,small,sf},labelfont=bf,labelsep=space]{caption}
\usepackage{float}
\usepackage{fancyhdr}
\usepackage{fnpos}
\usepackage[english]{babel}
\usepackage{array}
\usepackage{droidsans}
\usepackage{charter}
\usepackage[T1]{fontenc}
\usepackage[usenames,dvipsnames]{xcolor}
\usepackage{setspace}
\usepackage[compact]{titlesec}

\usepackage{epstopdf}

\definecolor{cream}{RGB}{222,217,201}

\begin{document}

\pagestyle{fancy}
\thispagestyle{plain}
\fancypagestyle{plain}{

}

\makeFNbottom
\makeatletter
\renewcommand\LARGE{\@setfontsize\LARGE{15pt}{17}}
\renewcommand\Large{\@setfontsize\Large{12pt}{14}}
\renewcommand\large{\@setfontsize\large{10pt}{12}}
\renewcommand\footnotesize{\@setfontsize\footnotesize{7pt}{10}}
\makeatother

\renewcommand{\thefootnote}{\fnsymbol{footnote}}
\renewcommand\footnoterule{\vspace*{1pt}%
\color{cream}\hrule width 3.5in height 0.4pt \color{black}\vspace*{5pt}} 
\setcounter{secnumdepth}{5}

\makeatletter 
\renewcommand\@biblabel[1]{#1}            
\renewcommand\@makefntext[1]%
{\noindent\makebox[0pt][r]{\@thefnmark\,}#1}
\makeatother 
\renewcommand{\figurename}{\small{Fig.}~}
\sectionfont{\sffamily\Large}
\subsectionfont{\normalsize}
\subsubsectionfont{\bf}
\setstretch{1.125} 
\setlength{\skip\footins}{0.8cm}
\setlength{\footnotesep}{0.25cm}
\setlength{\jot}{10pt}
\titlespacing*{\section}{0pt}{4pt}{4pt}
\titlespacing*{\subsection}{0pt}{15pt}{1pt}

\fancyfoot{}
\fancyhead{}
\renewcommand{\headrulewidth}{0pt} 
\renewcommand{\footrulewidth}{0pt}
\setlength{\arrayrulewidth}{1pt}
\setlength{\columnsep}{6.5mm}
\setlength\bibsep{1pt}

\makeatletter 
\newlength{\figrulesep} 
\setlength{\figrulesep}{0.5\textfloatsep} 

\newcommand{\topfigrule}{\vspace*{-1pt}%
\noindent{\color{cream}\rule[-\figrulesep]{\columnwidth}{1.5pt}} }

\newcommand{\botfigrule}{\vspace*{-2pt}%
\noindent{\color{cream}\rule[\figrulesep]{\columnwidth}{1.5pt}} }

\newcommand{\dblfigrule}{\vspace*{-1pt}%
\noindent{\color{cream}\rule[-\figrulesep]{\textwidth}{1.5pt}} }

\makeatother

\twocolumn[
  \begin{@twocolumnfalse}
\sffamily
%
%
%

\noindent\LARGE{\textbf{Electronic structure and time-dependent description of rotational predissociation of LiH}} \\
\quad \\
\noindent\large{P. Jasik,\textit{$^{a}$} J. E. Sienkiewicz, \textit{$^{a\ast}$} J. Domsta \textit{$^{b}$} and N. E. Henriksen\textit{$^{c}$}} \\
\quad \\
\noindent\normalsize{Adiabatic potential energy curves of the $^1\Sigma^+$ and $^1\Pi$ states of the LiH molecule have been  calculated. They correlate asymptotically to atomic states, like 2s+1s, 2p+1s, 3s+1s, 3p+1s, 3d+1s, 4s+1s, 4p+1s and 4d+1s. Very good agreement is found between our calculated spectroscopic parameters and experimental ones.  The dynamics of the rotational
predissociation process of the $1^1\Pi$ state has been studied by solving the time-dependent
Schr\"{o}dinger equation. The classical experiment of Velasco [Can. J. Phys. {35}, 1204 (1957)] on dissociation in  the $1^1\Pi$ state is explained in detail. } \\

 \end{@twocolumnfalse} \vspace{0.6cm}

  ]

\renewcommand*\rmdefault{bch}\normalfont\upshape
\rmfamily
\section*{}
\vspace{-1cm}

\footnotetext{\textit{$^{a}$~Faculty of Applied Physics and Mathematics, Gda\'{n}sk University of Technology, Narutowicza 11/12, 80-233 Gda\'{n}sk, Poland}}
\footnotetext{\textit{$^{\ast}$~E-mail: jes@mif.pg.gda.pl}}
\footnotetext{\textit{$^{b}$~Institute of Applied Computer Sciences, State University of Applied Sciences in Elblag, Wojska Polskiego 1, 82-300 Elblag, Poland}}
\footnotetext{\textit{$^{c}$~Department of Chemistry, Building 207, Technical
University of Denmark, DK-2800 Kgs. Lyngby, Denmark.}}




\section{INTRODUCTION}\label{INTRODUCTION}
During the last twenty years, the physics of diluted gases has seen
major advances in two fields, namely laser cooling of atomic and
molecular samples and femtosecond chemistry. In both cases,
appropriately frequency and phase shaped laser light is used in
order to control the system. In this context, two fundamental
processes, i.e., photoassociation and photodissociation, or in other
words formation and breaking of the chemical bond by light, have
attracted attention of theoreticians as well as
experimentalists. Particularly, photodissociation of diatomic or
small polyatomic molecules is an ideal field for investigating
molecular dynamics at a high level of precision.

Homonuclear and heteronuclear alkali metal molecules, including LiH, are valuable for  theoreticians, mainly because they have simple electronic
structure, being one- or two-valence electron systems. They can
serve as convenient prototypes to test theoretical methods, which
can be further applied to more complicated molecular systems.
Besides that, the knowledge of interatomic adiabatic potential
energy curves of diatomic systems is essential in understanding
several
 processes like photodissociation, photoassociation, cooling and
 trapping. An extensive  survey on the spectroscopy and structure
 of LiH was published in 1993 by Stwalley and Zemke
 \cite{Stwalley1993};  and later on they were followed by Gadea
 \cite{Gadea2006} in 2006.

Already in 1935-1936, Crawford and Jorgensen
\cite{Crawford1935, Crawford1936} made analysis of the LiH band spectra.
Since that, many notable studies have been undertaken. Among them,
in 1962 Singh and Jain \cite{Singh1962} applied the
Rydberg-Klein-Rees method in order to obtain energies of the low
excited states of LiH. Gadea and coworkers calculated potential
energy curves \cite{Boatlib1992, Casida2000, Gadea2006, 2016Beriche}, radial
couplings \cite{Gadea1993}, nonadiabatic energy shifts
\cite{Gemperle1999} as well as the LiH formation by radiative
association in ion collisions \cite{Dickinson2000}. Results of
several other calculations, including semiempirical and ab initio
approaches to describe important physical and chemical properties of
LiH are available
 \cite{1982Roos, 2000Sharma, Gianturco1996a, Gianturco1996b, Bubin2004, Stancil1997,
 Bodo2003, Fondermann2007, Trail2008, Aymar2009, 2009Cooper}.
Calculations related to LiH are also used in the description of
ultracold
 polar molecules formation in a single quantum state (e.g. Cote et al.
 \cite{Cote2010}). Special investigations were
 devoted to dipole moments \cite{2002Cafiero, 2009Fernandez} and ionic states of LiH
 \cite{Decleva1986, 2004Magnier, 2007Cheng}. Recently, Tung et
al. \cite{2011Tung} and Holka et al. \cite{2011Holka} performed very
accurate calculations of the ground and some excited state potential curves.

 LiH was also intensively explored in time-dependent studies.
 Again, being only a four electron molecule makes it a convenient
example for molecular dynamics calculations. Already in 1936
Mulliken \cite{1936Mulliken} noted that the change in the
internuclear separation may cause a rearrangement of the
density of electrons' distribution. Recently, the LiH molecule was
used in a computational study using the time-dependent multiconfiguration
method \cite{2008Nest}.

The aim of our work is to provide accurate potential energy curves and next to use them to explain a classical experiment of Velasco \cite{1957Velasco} on rotational predissociation. We choose to solve the time dependent Schr\"odinger equation (TDSE) with a probe wavepacket placed on the effective interatomic potential possessing a   centrifugal barrier. This approach gives us the possibility to compare rovibrational spacings with those calculated directly from the electronic structure.  Our work is also motivated by the case of the NaI molecule intensively studied by A. Zewail \cite{1994Zewail} and later by others \cite{1996Gronager, 1998Gronager, 1998Dietz,
2000Moller}. The NaI dimer shows similar behavior as LiH in creating ionic bonds and is a well studied prototype molecule in femtochemistry, particularly in the aspect of dynamics of unimolecular
reactions.

 In Section II,
the appropriate model of the electronic structure is defined,
leading to an algorithm for calculating some low-excited singlet
$\Sigma$ and $\Pi$ states. Next, we describe the theoretical  backgrounds of rotational predissociation and molecular dynamics. We explain, how the obtained
adiabatic potentials can be  used in the theoretical treatment of
the rotational predissociation proccess. In Section III, we present rotational
predissociation results for the $1^1\Pi$ state and compare them with measurements
of Velasco \cite{1957Velasco}. 
Finally, we show results of the
dynamics of predissociation process induced by a laser field. Conclusions are given in the last section.

\section{THE MODEL}\label{MODEL}

\subsection{Electronic structure}

We consider the interaction between the lithium (atom A) and hydrogen (atom B) under the assumption that the molecular state is a composition of
the electronic adiabatic states $\Psi_{i}^{el}(\vec{r};{R})$, $i =
1,2,3,\dots$, which depend on the positive variable $R$, i.e. on the
separation between the nuclei of these atoms. The applied  notation indicates, that our considerations are restricted to such eigenstates,
which are  independent of the direction of the vector joining the
nuclei. In other words, electronic wave functions possess the spherical
symmetry with respect to the nuclear coordinates. Our calculations are based on the Born-Oppenheimer
approximation, i.e. as the solutions of the following time
independent Schr\"{o}dinger equation
\begin{eqnarray}\label{SE}
H^{el} \Psi_{i}^{el}(\vec{r};{R}) &=&
E_{i}^{el}(R)\Psi_{i}^{el}(\vec{r};{R}).
\end{eqnarray}
Here, the separation parameter $R$ is kept fixed, vector $\vec{r}$
represents all electronic coordinates, $H^{el}$ is the electronic
Hamiltonian of a diatomic system. Thus  $\Psi_{i}^{el}(\vec{r};{R})$
describes the $i$-th eigenstate of the Hamiltonian, $E_{i}^{el}(R)$
are the corresponding eigenvalues, also named as adiabatic potentials.
The Hamiltonian of the system can be written as
\begin{eqnarray}\label{Hel}
H^{el} &=& T^{el} + V,
\end{eqnarray}
where $T^{el}$ stands for the kinetic energy operator of the valence
electrons and $V$ represents the operator of the interaction between
the valence electrons, the Li-core and the nucleus of H.
In the present
approach only the valence electrons are treated explicitly and the
lithium core is  represented by an angular momentum dependent
pseudopotential. 
 The latter is taken as
\begin{eqnarray}
V &=& V^{A} + V^{A}_{pol} + V^B + \frac{1}{r_{12}}
+ V_{cc}.
\end{eqnarray}
Here $V^{A}$ describes Coulomb and exchange interaction as well as
the Pauli repulsion between the valence electrons and the lithium
core. We use the following semi-local energy-consistent
pseudopotentials:
\begin{eqnarray}
  V^{A} &=& \sum_{i=1}^{2}\bigg{(}-\frac{Q_{A}}{r_{A i}} +
  \sum_{l, k}B^{A}_{l, k}\exp(-\beta^{A}_{l, k}r_{A i}^{2})P^{A}_{l}\bigg{)},
\end{eqnarray}
where $Q_{A} = 1$ denotes the net charge of the lithium core,
$P^{A}_{l}$ is the projection operator onto the Hilbert subspace of
angular symmetry $l$ with respect to the Li$^{+}$-core. The parameters $B^{A}_{l, k}$
and $\beta^{A}_{l, k}$ define the semi-local energy-consistent
pseudopotential. The second interaction term in Eq. (3) is the
polarization term which describes, among others, core-valence
correlation effects and is taken as
\begin{eqnarray}
V^{A}_{pol} &=& -\frac{1}{2} \alpha_{A} \vec{F}^{2}_{A},
\end{eqnarray}
where $\alpha_{A} =$ 0.1915 $a_{0}$  is the dipole polarizability of
the lithium core \cite{1982Fuentealba} and $\vec{F}_{A}$ is the
electric field generated at its site by the valence electrons. For the latter we are using the following formula 
\begin{eqnarray}
\vec{F}_{A} &=& \sum_{i} \frac{\vec{r}_{A i}}{r^{3}_{A i}} [1 -
\exp(-\delta_{A} r^{2}_{A i})],
\end{eqnarray}
where $\delta_{A}$ is the cutoff parameter, which equals
 0.831 $a_{0}^{-2}$ (value taken from Fuentealba et al.\cite{1982Fuentealba}). The third term in Eq. (3)
represents the Coulomb interaction between the valence electrons and
the hydrogen nucleus. The fourth term stands for the repulsion
between the valence electrons, whereas the last term describes the
interaction between the lithium core  and hydrogen nucleus. Since the lithium
atomic core and the hydrogen nucleus  are well separated, we choose
a simple point-charge Coulomb interaction in the latter case. More
detailed characteristics of the applied potentials  are given in the
papers of Czuchaj and co-workers \cite{Czuchaj1998,Czuchaj2003a} and
Dolg \cite{Dolg2000}.

 The core electrons of the Li
atom are represented by the pseudopotential ECP2SDF
\cite{1982Fuentealba}, which was formed from the uncontracted
($9s9p8d3f$) basis set. The basis for the $s$ and $p$ orbitals,
which comes with this potential is enlarged by functions for $d$ and
$f$ orbitals given by P. Feller \cite{2006Feller} and assigned by
cc-pV5Z.  Additionally, our basis set was augmented by four $s$ short
range correlation functions (1979.970927, 392.169555, 77.676373,
15.385230), four $p$ functions (470.456384, 96.625417, 19.845562,
4.076012), four $d$ functions (7.115763, 3.751948, 1.978298, 1.043103)
and four $f$ functions (2.242072, 1.409302, 0.885847, 0.556818). Also,
we added to the basis the following set of diffuse functions: two $s$ functions
(0.010159, 0.003894), two $p$ functions (0.007058, 0.002598), two $d$
functions (0.026579, 0.011581) and two $f$ functions (0.055000,
0.027500). The numbers in parenthesis are coefficients of the
exponents of the primitive Gaussian orbitals. The basis set for the
hydrogen electron is the standard cc-pV5Z basis \cite{2006Feller}.

The spin-orbit coupling (SO)
contributes insignificantly to the energy of our system, so we do
not take it into account.
\begin{table*}
\caption{\ Comparison of asymptotic energies with other theoretical and experimental results. Energies are shown in cm$^{-1}$ units\\[0.3cm]}
\label{tbl:asymptots}
\begin{tabular*}{\textwidth}
{@{\extracolsep{\fill}}ccccccc}
  \hline
  Atomic asymptotes & Experiment Moore \cite{1949Moore} & Theory Boutalib \cite{Boatlib1992} & Theory Gadea \cite{Gadea2006} & Theory present \\
  \hline
  Li(2p)+H(1s) & 14904 & 14905 & 14898 & 14904 \\
  Li(3s)+H(1s) & 27206 & 27210 & 27202 & 27202 \\
  Li(3p)+H(1s) & 30925 & 30926 & 30920 & 30921 \\
  Li(3d)+H(1s) & 31283 & 31289 & 31279 & 31276 \\
  Li(4s)+H(1s) & 35012 & 35018 & 35007 & 35016 \\
  Li(4p)+H(1s) & 36470 & 36475 & 36465 & 36464 \\
  Li(4d)+H(1s) & 36623 & 37590 & 36626 & 36617 \\
  \hline
\end{tabular*}
\end{table*}
To calculate adiabatic potential energy curves of the LiH diatomic
molecule we use the  multiconfigurational self-consistent
field/complete active space self-consistent field (MCSCF/CASSCF)
method and the multi-reference configuration interaction (MRCI)
method. All calculations are performed by means of the MOLPRO
program package \cite{MOLPRO2006}. Using these computational methods
we obtained adiabatic potential energy curves for singlet $\Sigma$, $\Pi$ and $\Delta$ states, which correlate to the
Li(2s)+H(1s) ground atomic asymptote and the Li(2p)+H(1s),
Li(3s)+H(1s), Li(3p)+H(1s), Li(3d)+H(1s) excited atomic asymptotes,
respectively. The quality of our calculations can be confirmed by
the comparison with experimental and theoretical asymptotic energies
for different electronic states, which is shown in Table 1. 
Our
asymptotic energies for ground and excited states are in very good
agreement with experimental and other theoretical values. Particularly, perfect match is found between our result and the experimental value for the Li(2p) energy level.  

\subsection{Rotational predissociation}

When the adiabatic potential $E^{el}(R)$ of the singlet state
$^1\Lambda$ is obtained from solution of Eq. (\ref{SE}), the
effective potential energy may be written in the following form
(e.g. Landau and Lifshitz) \cite{1965Landau}:

\begin{equation}\label{UJ1}
U_J(R) = E^{el}(R) + \frac{J(J+1) - \Lambda^2 }{2\mu
R^2},
\end{equation}
where $\Lambda$ is the component of the sum over all electron angular momenta on the
diatomic axis, $J \ge \Lambda$ is the rotational quantum
number of the molecule,   and $\mu$ is the reduced mass of the nuclei.

 Rovibrational energies $E(v,J)$ depend on $E^{el}(R)$ as well as vibrational
$v$ and rotational  $J$ quantum numbers. They
 are solutions of the time-independent nuclear Schr\"{o}dinger
equation:

\begin{equation}\label{ntSE}
H^{nuc} \Psi^{nuc}_{v,J}(R) = E(v,J) \Psi^{nuc}_{v,J}(R),
\end{equation}
where the nuclear Hamiltonian is taken as

\begin{equation}\label{Hmol}
H^{nuc} = -\frac{\hbar^2}{2\mu}\frac{\partial^2}{\partial R^2} +
U_J(R).
\end{equation}

The effective potential $U_J(R)$ forms a  barrier for $J>0$ with a  maximum
$U_J(R_J)$, at the internuclear distance $R_J$, which easily can be
estimated. Any rovibrational state with the positive energy $E(v,J)$
lower than $U_J(R_J)$ has a finite lifetime before it will be
decomposed due to a quantum  tunneling effect. These states are called
quasibound states and formally  belong to the continuum.  What is
important is that during their lifetimes they can be regarded as bound
states. When the energy $E(v,J)$ exceeds the barrier maximum
$U_J(R_J)$ then any bound state is not possible. Following Way and
Stwalley \cite{1973Way}, we introduce a critical value of
the rotational quantum  number $J_c$ which obeys the two following
inequalities:

\begin{equation}\label{bound)}
E(v,J_c) < U_{J_c}(R_{J_c})
\end{equation}

and

\begin{equation}\label{cont)}
E(v,J_c + 1) > U_{J_c + 1}(R_{J_c + 1}).
\end{equation}

In other words, for a given $v$, the state with the energy
$E(v,J_c)$ is the last of the quasibound states series  supported by
the barrier, and the state with the energy $E(v,J_c + 1)$ already
belongs to the continuum. By solving Eq. \ref{ntSE} we obtain
$E(v,J_c)$ and estimate by extrapolation $E(v,J_c + 1)$.
Respectively, the differences  $E(v,J_c) - E(0,0)$ and $E(v,J_c+1) -
E(0,0)$ may refer to the last observed and the first unobserved
rotational predissociation experimental result.

\subsection{Molecular dynamics}

The time-dependent approach which is mathematically equivalent to
the time-independent one can be regarded as a complimentary tool and is often used in studying photodissociation processes.
Here, it serves as an alternative and quite illustrative method for
testing results of our structural calculations.

We start our consideration from the time-dependent Schr\"{o}dinger
equation written in the following form
\\
\begin{equation}\label{TDSE}
\imath \hbar \frac{\partial}{\partial t} \Phi(R,t) = H^{nuc}
\Phi(R,t),
\end{equation}
where $\Phi(R,t)$ is the time dependent wavepacket moving on the
effective potential energy curve $U_J(R)$ (Eq. \ref{UJ1}) and
$H^{nuc}$ is the nuclear Hamiltonian given in Eq. 9.

By definition the wave-packet is a coherent superposition of
stationary states (e.g. Tannor
\cite{2007Tannor}) which may be represented in the following
form consisting of two contributions from the discrete and  continuous parts of the spectrum
\begin{equation}\label{pakiet falowy}
\Phi({R}; t) = \sum_{v,J} c_{v,J} \, \Psi^{nuc}_{v,J}(R) \,  \mathrm{e}^{-\imath E(v,J)t / \hbar}
 +  \int  
\, c(E) \, \Psi_{E}(R)\, \mathrm{e}^{-\imath Et / \hbar} \, dE,
\end{equation}
where $c_{v,J}$ and $c(E)$ are the energy-dependent coefficients, $\mathrm{e}^{-\imath E(v,J)t / \hbar}$ and 
$\mathrm{e}^{-\imath Et / \hbar}$ are the time evolution factors, 
$\Psi^{nuc}_{v,J}(R)$ and $\Psi_{E}(R)$ are eigenfunctions of ${H}^{nuc}(R)$. The
wavepacket $\Phi(R; t)$ is a solution of Eq. (\ref{TDSE}) and
its initial shape at $t = 0$ may be  taken as a Gaussian function of
arbitrary half-width placed on the effective potential energy curve. The
wavepacket moves away from its starting location due to the Newtonian force
$-dU_J/dR$. This process is described by the time-dependent
autocorrelation function

\begin{equation}\label{funkcja autokorelacji}
S(t) =  \int \Phi({R}; t=0) \,  \, \Phi({R}; t)
\, dR.
\end{equation}
\\

In our case
the autocorrelation function describes evolution of the initial wave
packet in the excited electronic state. The time-dependent wavepacket population is calculated as
\begin{equation}\label{population_J}
P(t) =  \int_0^{R_{max}} | \Phi({R}; t) |^2\, dR.
\end{equation}
The expression for the
absorption cross section is proportional to the Fourier transform of
$S(t)$ and is written as:
\\
\begin{equation}\label{tacs}
\sigma(E) = \frac{E}{2\hbar^2 c} \, 
\int_{-\infty}^{+\infty} \mathrm{e}^{\imath E t / \hbar} \, S(t) \,
dt,
\end{equation}
\\
where $E$ is the photon energy.

\section{RESULTS AND DISCUSSION}


Our results of the calculated adiabatic potential curves of $1-8^1
\Sigma^+$ and $1^1\Pi$ states are presented in FIG. 1.
\begin{figure}[ht]
\centering
\includegraphics[height=7cm]{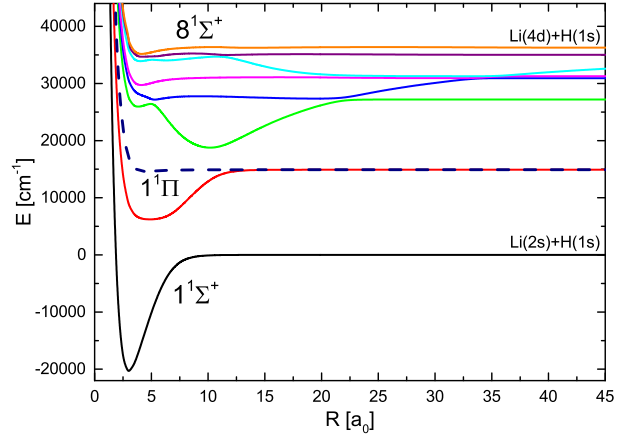}
\caption{Adiabatic potential energy curves of LiH:
1-8 $^1\Sigma^+$ states (solid lines), 1$^1\Pi$ state (dashed 
line).}
\label{singlety}
\end{figure}
Several characteristic avoided crossings are visible, particularly the double one at 5 and 20 $a_{0}$ between the curves of the $3^1\Sigma$
and $4^1\Sigma$ states. Though not very pronounced, there are
avoided crossings between $1^1\Sigma$ and $2^1\Sigma$ at 7.5 $a_{0}$
and  $2^1\Sigma$ and $3^1\Sigma$ at 10 $a_{0}$. 

Equilibrium
positions R$_e$ and depths of the potential wells D$_e$ are compared
with other theoretical and experimental results in Table 2.
\begin{table*}[ht]
  \caption {\ Spectroscopic parameters R$_{e}$ [a$_0$], D$_{e}$, $\omega_e$, and T$_e$ [cm$^{-1}$] for the ground and low-excited states of the $\rm LiH$ molecule\\[0.3cm]}
  \label{tbl:parameters}
  \begin{tabular*}{\textwidth}
  {@{\extracolsep{\fill}}l c l c c c c}
  \hline
State & Dissociation limit & Author & R$_e$ & D$_e$ & $\omega_e$ & T$_e$ \\
  \hline
$1^1 \Sigma^+$ & Li(2s) + H(1s) & present (theory) & 3.003 & 20327 & 1391 &  \\
 &  & Aymar 2009 (theory) \cite{Aymar2009} & 3.002 & 20167 & 1398 & \\
 &  & Gadea 2006 (theory) \cite{Gadea2006} & 3.003 & 20349 &  & \\
 &  & Dolg 1996 (theory) \cite{1996Dolg} & 3.000 & 20123 & 1391 & \\
 &  & Stwalley 1993 (exp.) \cite{Stwalley1993} & 3.015 & 20288 & 1407 & \\
 &  & Boutalib 1992 (theory) \cite{Boatlib1992} & 3.007 & 20174 &  &  \\
 &  &  &  &  &  & \\
$2^1 \Sigma^+$ & Li(2p) + H(1s) & present (theory) & 4.866 & 8687 & 260 & 26544 \\
 &  & Aymar 2009 (theory) \cite{Aymar2009} & 4.820 & 8698 & 241 & \\
 &  & Gadea 2006 (theory) \cite{Gadea2006} & 4.862 & 8687 &  & \\
 &  & Stwalley 1993 (exp.) \cite{Stwalley1993} & 4.906 & 8679 &  &  \\
 &  & Boutalib 1992 (theory) \cite{Boatlib1992} & 4.847 & 8690 &  & 26390 \\
 &  & Vidal 1982 (theory) \cite{1982Vidal} & 4.910 & 8686 & 244 & \\
 &  &  &  &  &  & \\
$1^1 \Pi$ &  & present (theory) & 4.50 & 286 & 226 & 34945 \\
 &  & Velasco 1957 (exp.) \cite{1957Velasco} & 4.49 & 284 & 216 & \\
 &  & Vidal 1982 (theory) \cite{1982Vidal} & 4.50 & 289 &  & \\
 &  & Aymar 2009 (theory) \cite{Aymar2009} & 4.52 & 251 & 243 & \\
 &  &  &  &  &  & \\
$3^1 \Sigma^+$ & Li(3s)+H(1s) & present (theory) & 3.821 & 1270 & 540 & 46259 \\
 &  &  & 10.172 & 8438 & 293 & 39092 \\
 &  & Aymar 2009 (theory) \cite{Aymar2009} & 3.830 & 1267 & 390 & \\
 &  &  & 10.150 & 8361 & 390 & \\
 &  & Gadea 2006 (theory) \cite{Gadea2006} & 3.821 & - &  & \\
 &  &  & 10.181 & 8453 &  & \\
 &  & Huang 2000 (exp.) \cite{Huang2000} & - & - &  & \\
 &  &  & 10.140 & 8469 &  & \\
 &  & Boutalib 1992 (theory) \cite{Boatlib1992} & 3.825 & 1277 &  & 46109 \\
 &  &  & 10.206 & 8444 &  & 38942 \\
\hline
\end{tabular*}
\end{table*}
For the ground state our position of $R_e$ agrees exactly with the
theoretical value of Dolg \cite{1996Dolg} and reasonably with the
experimental value of Stwalley et al. \cite{Stwalley1993}. We also
find a good agreement within 40 cm$^{-1}$ between the well depths $D_e$ of our results and experimental data of Stwalley et al. In the case of 1$^1\Pi$, our results of R$_e$ and D$_e$ agree within 2 cm$^{-1}$ with the experimental data of Velasco. All theoretical
results indicate the existence of a double well for the $3^1\Sigma$
state but this is not confirmed by the only available experiment by
Huang et al. \cite{Huang2000}.

FIG. 2 displays spacings between successive rovibrational levels
of the $1^1\Pi$ state. Our first set of values is  obtained by solving \cite{LeRoy2011} Eq. \ref{ntSE}. The second one comes from appropriate differences
between the positions of peaks in the absorption spectrum obtained
from Eq. \ref{tacs} and presented  in FIG. \ref{spec_final}.
These two sets agree very well with each other. Moreover,  there is
also very good agreement with the experimental values of Velasco
\cite{1957Velasco}.
\begin{figure}[H]
\centering
\includegraphics[height=7cm]{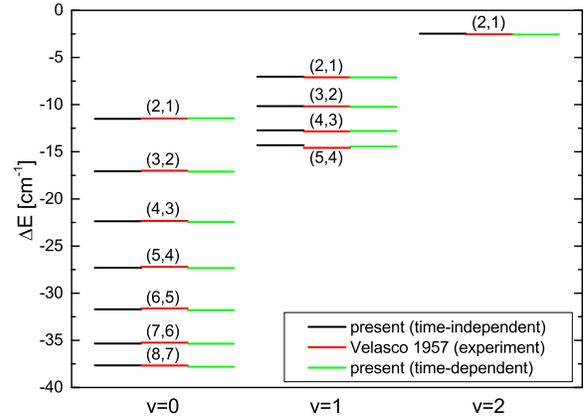}
\caption{Differences $\Delta E(v, J', J)=
E(v,J')-E(v,J)$ between rovibrational levels with the same
vibrational quantum number $v$ of the $1^1\Pi$ state. Three series
of differences are drawn for $v=1, 2$ and $3$. Each difference is
specified by $(J,J')$. The blue lines are coming from
calculated rovibrational levels. The red ones are derived from
experimental data of Velasco \cite{1957Velasco}. The green ones are
our results obtained from the absorption spectrum shown in FIG. 3.}
\label{rot_levels1}
\end{figure}

\begin{figure}[H]
\centering
\includegraphics[height=7cm]{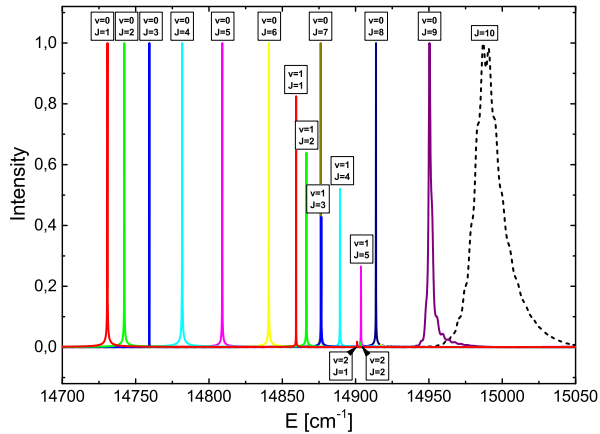}
\caption{The total absorption cross
section calculated from Eq. \ref{tacs}.}
\label{spec_final}
\end{figure}
The peaks in the absorption spectrum (FIG. 3) are obtained by solving the time dependent Schr\"{o}dinger equation\cite{Schmidt2009} (Eq. \ref{TDSE})  with a Gaussian-shaped wavepacket $\Phi$ initially centered at 6.15 $a_0$ and possessing the half-width equal to $0.95\ a_0$. Here, we are not interested in the intensity of the peaks and the precise shape of the initial wave packet is unimportant. The set of effective potentials $U_J$ (Eq. \ref{UJ1}) spans $J$ from $1$ to $10$. The broadened peak labeled by $v=0$ and $J=9$ is the last in the series since $J=9$ is a critical value $J_c$ discussed in Section 2.2. Its half-width (FWHM) is equal to 2.7 cm$^{-1}$. The last very broad peak with $J=10$ illustrates the situation where the depth of the effective potential is too shallow to allow for existence of any bound vibrational level. The last and already broadened peak observed by Velasco was assigned as $v=0$ and $J=8$. In his analysis, he correctly foreseen  the existence of an unobserved peak labeled by $v=0$ and $J=9$ before the molecule breaks off due to high rotations.  But his prediction of existence of two other missing peaks in the spectrum, namely with $v=1$, $J=6$ and $v=2$, $J=3$ is not confirmed by our results. The broadening of the peak with $v=0$ and $J=9$ showed by our calculation is due to quantum tunneling through the centrifugal barrier.
\begin{figure}
\centering
\includegraphics[height=7cm]{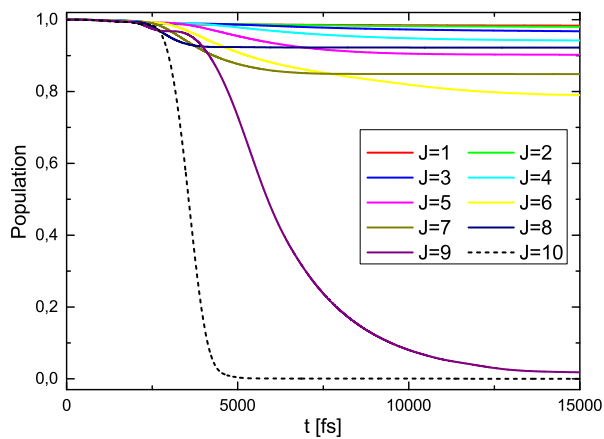}
\caption{Time-dependent population of the
wavepacket placed on the effective potential $U_J(R)$ (J=1,...,10) for electronic energy of the $1^1\Pi$ state. All lines refer to the same initial conditions at $t=0$ of the wavepacket.}
\label{population}
\end{figure}

The last figure (FIG. 4) shows the results for the time dependent population of the
1$^1\Pi$ state for the same initial condition. For $J=10$, no bound states are supported
by the effective potential and the drop in population around 2.5 ps shows the time
it takes for the continuum wave packet to reach $R = R_{max}$. In our calculations, we set this value to be equal to 100  $a_0$. In all cases, the
population is close to one within the first approximately 2.5 ps, since any
continuum part of the wave packet needs this time to reach $R_{max}$. Furthermore for
low values of $J$, the population is close to one within the time window of 15 ps,
meaning that essentially all parts of the wave packet can be represented by bound
states. For $J=9$, the wave packet consists of a continuum as well as a (quasi-) bound
part. The quasibound part decays via tunneling giving rise to the slow
exponential decay with a decay constant of 2.4 ps. 
Based on the time-energy uncertainty principle, we can estimate that this lifetime
should give rise to a line width of
approximately 2 cm$^{-1}$. This is in good agreement with the spectrum in FIG. 3.


\section{Conclusions}

In order to describe the rotational predissociation  process of the
LiH molecule we start from calculating the low lying adiabatic
potential energy curves with particular emphasis on the $1^1\Pi$ state. Our spectroscopic parameters are in very good agreement with experimental values.
Having the potential curve of $1^1\Pi$ state we calculate the
rovibrational levels. The differences between these successive
levels are compared with those derived from experimental data of
Velasco. The agreement again is very good, which means that the shape of
the first excited electronic state $1^1\Pi$ is reliable. On the
other hand since our difference ($T_e$) between potential
wells of $1^1\Pi$  and of the ground state $1^1\Sigma^+$ is around
50 cm$^{-1}$ larger than experimental value of Stwalley et al., the
direct comparison with  the spectrum of Velasco shows a small systematic
shift. 

In order to get insights from the complementary time-dependent
approach we solve the time-dependent nuclear Schr\"{o}dinger
equation. The solution shows the evolving wavepacket originally
placed on the effective potential curve. The absorption spectrum is
calculated as a Fourier transform of the autocorrelation function. The
differences between  successive peaks in the spectrum are compared
with those of Velasco and ours obtained in the time-independent
approach. All three sets of values are in very good agreement. Our results for time-dependent population of the $1^1\Pi$ state  explain in detail the rotational  predissociation mechanism of the LiH molecule. A challenge for experimentalist would be to detect in real time (via pump-probe spectroscopy) the predissociation due to quantum tunneling through the centrifugal barrier.

\section*{Acknowledgments}
This work was partially supported by the COST action XLIC (CM1204) of the
European Community. Calculations have been carried out using resources of the Academic Computer Centre in Gda\'nsk.




\bibliographystyle{rsc} 

\end{document}